# Three-dimensional solitary waves with electrically tunable direction of propagation in nematics

Bing-Xiang Li,[1, 2] Rui-Lin Xiao,[1, 2] Sathyanarayana Paladugu,[1] Sergij V. Shiyanovskii,[1, 2] and Oleg D. Lavrentovich[1, 2, 3*]

[1]*Advanced Materials and Liquid Crystal Institute, Kent State University, Kent, Ohio, 44242, USA*

[2]*Chemical Physics Interdisciplinary Program, Kent State University, Kent, Ohio, 44242, USA*

[3]*Department of Physics, Kent State University, Kent, Ohio, 44242, USA*

*olavrent@kent.edu

Production of stable multidimensional solitary waves is a grand challenge in modern science[1]. Steering their propagation is an even harder problem. In this work we demonstrate three-dimensional solitary waves in a nematic, trajectories of which can be steered by the electric field in a plane perpendicular to the field. The steering does not modify the properties of the ground state that remains uniform. These localized waves, called director bullets, are topologically unprotected multidimensional solitons of (3+2)D type that show fore-aft and right-left asymmetry with respect to the background molecular director; the symmetry is controlled by the field. Besides bringing a fundamental novelty by adding a whole dimension to the propagation direction, the solitons can lead to applications such as targeted delivery of information and micro-cargo.

Solitary waves maintain their self-confined shapes while propagating and surviving collisions with each other. A classic example is a one-dimensional wave in a shallow water channel observed by Russell [2,3]. Creation of solitons of higher dimensions represents a major challenge in the science of nonlinear fields and matter [1]. The multidimensional solitons are abbreviated as (*m*+1)D objects, where "1" represents the propagation direction and "*m*" shows in how many dimensions the soliton is self-focused [4,5]. The Russell's wave is (1+1)D type, as it propagates along the channel and water displacement occurs along a single transverse direction. Multidimensional solitons are of practical importance [1], for example, in optics, where the so-called "light bullets" of a (3+1)D type could serve as information carriers [6]. Another grand challenge of soliton research is whether and how their direction of propagation could be guided [4]. In a uniform medium, solitons propagate along rectilinear trajectories. To curve these trajectories, one could use interactions of solitons with other waves [5], with each other [7], or by designing a spatially-varying medium. One example are nematicons, (2+1)D light beams propagating in a nematic [1,4]. A nematic shows a uniaxial molecular orientation along the so-called director $\hat{\mathbf{n}}$. The nematicons can be bent by a spatially-varying director $\hat{\mathbf{n}}(\mathbf{r})$ [8].

In this work, we present experimental realization of multidimensional solitary waves of a (3+2)D type. These waves of director deformations in a nematic are self-localized in three spatial dimensions, with the propagation direction controlled by the applied electric field in the plane perpendicular to the field. To the best of our knowledge, there are no other examples of stable and steerable (3+2)D solitons, neither in matter nor in a field.

We used a nematic 4'-butyl-4-heptyl-bicyclohexyl-4-carbonitrile (CCN-47) doped with 0.005wt% of ionic salt tetrabutylammonium bromide (TBABr). The material is of the (-,-) type, with negative dielectric $\Delta\varepsilon = -3.3$ and conductivity $\Delta\sigma = -2\times10^{-9}\ \Omega^{-1}\mathrm{m}^{-1}$ anisotropy. The cell of thickness $d = 8.0\ \mu\mathrm{m}$ is composed of two glass substrates coated with transparent electrodes and surface alignment layers to set planar orientation $\hat{\mathbf{n}}_0 = (0,1,0)$ in the *xy* plane parallel to the bounding plates. A sinusoidal AC electric field of frequency $f = 1\text{-}10^3$ Hz is applied normal to the substrates, $\mathbf{E} = (0,0,E)$.

In the range 5 Hz$< f <$27 Hz, the electric field produces a new type of solitary waves of director deformations, which are (3+2)D director bullets abbreviated as $B_\alpha^1$; the superscript means "low frequency" and the subscript indicates the angle between $\mathbf{v}$ and $\hat{\mathbf{n}}_0$, Figs. 1, 2, Supplementary Fig. 1. At $U = const$, $B_\alpha^1$ solitons move either parallel to $\hat{\mathbf{n}}_0$, Fig. 1, or perpendicularly to it, Fig. 2, Supplementary Fig. 2. We call these $B_0^1$ and $B_{90}^1$ bullets, respectively. The stability range of $B_\alpha^1$ is limited by frequency, $5\,\text{Hz} \leq f \leq 27\,\text{Hz}$, and voltage, $4\,\text{V} \leq U \leq 12\,\text{V}$, Fig. 3a. At $f = const$, the voltage increase produces first $B_0^1$ solitons, then coexisting $B_0^1$ and $B_{90}^1$, then solely $B_{90}^1$, Fig. 3b; the speed of solitons grows with $U$, Fig. 3c. At $U > U_{\text{EHD}}(f) \approx (12.8 + 0.08\,\text{s}\,f)\,\text{V}$, one observes periodic electrohydrodynamic (EHD) domains, Fig. 3a, Supplementary Fig. 3. At $f > 500\,\text{Hz}$, one observes dielectric bullets of (3+1)D type that propagate always perpendicularly to $\hat{\mathbf{n}}_0$ [9], Supplementary Fig. 4.

The most striking feature of the $B_\alpha^1$ bullets is that their trajectories and symmetry is controlled by the electric field. A step-like increase of $U$ from 8 V to 11 V at $f = const$, transforms $B_0^1$ into $B_{90}^1$, Fig. 3d, while a decrease of $U$ transforms $B_{90}^1$ into $B_0^1$, Fig. 3e. In both cases, the bullets change their propagation direction by $90°$, Fig. 3f; the trajectories can also be controlled by the voltage change rates, Figs. 3g, h. The solitons also exhibit an intriguing behavior near the limits of the stability islands. Once the voltage is increased above 12.5 V (20 Hz), the $B_{90}^1$ solitons stop. If the voltage remains fixed, they disappear within the decay time $\tau \approx$ 2-3 s that depends on the depth of voltage increase. However, if the voltage is reduced back within $\tau$, the solitons start to move, either in the same direction, or in the opposite direction, Fig. 3i, j. The outcome depends on the exact timing of voltage change. Similar effects of stoppage, disappearance, and reactivation are observed for $B_0^1$, when the voltage is reduced below 7.2 V, Supplementary Fig. 5 and Supplementary.

The $B_\alpha^1$ bullets reveal their soliton nature by surviving collisions, Fig. 4. Within different stability islands, solitons show different outcomes of collisions. If only one soliton type is stable, their collision results in reappearance of a similar pair. In Fig. 4a, b, two $B_0^1$ bullets move parallel to $\hat{\mathbf{n}}_0$ towards each other, coalesce and form a single perturbation that separates into two $B_{90}^1$ solitons that move perpendicularly to $\hat{\mathbf{n}}_0$. This symmetry change is only temporary, as the two reconstruct their $B_0^1$ structure and continue propagation parallel to $\hat{\mathbf{n}}_0$. When collisions happen in the phase diagram region where both types of solitons are stable, new channels of reaction emerge. For example, two $B_0^1$ solitons can merge into a single $B_{90}^1$ that propagates perpendicularly to the "parent" solitons, Fig. 4c,d. An opposite effect is also possible, with two $B_{90}^1$'s producing a single $B_0^1$. Collisions of $B_0^1$ and $B_{90}^1$ produce other scenarios, such as transformations $B_0^1 \rightarrow B_{90}^1$ and $B_{90}^1 \rightarrow B_0^1$.

To understand the unusual properties of the (3+2)D solitons, we explore their inner structure. The direction of propagation of the bullets is determined by the dynamic asymmetry of the in-plane director rotations $\varphi(x,y,z,t)$ away from $\hat{\mathbf{n}}_0$. Here $t$ is time. Because of surface anchoring, $\varphi = 0$ at the bounding plates, and reaches its maximum in the middle plane of the cell, $\varphi(x,y,z=d/2,t) = \varphi_m(x,y,t)$. In observations with two crossed polarizers, a non-zero $\varphi_m$ increases light transmittance through the sample. The polarizing optical microscope textures of $B_0^1$ and $B_{90}^1$ represent four quadrants of a nonzero $\varphi_m$, surrounded by a uniform ($\varphi=0$) background, Fig. 1a, 2a. The $B_0^1$ bullets propagating along $\hat{\mathbf{n}}_0$ lack the fore-aft symmetry, Fig. 1a. The $B_{90}^1$ bullets, propagating perpendicularly to $\hat{\mathbf{n}}_0$, lack the left-right symmetry, Fig. 2a.

To map the spatiotemporal variations of $\varphi_m$, we use two linear polarizers crossed at an angle 78° or 65° with each other. The light intensity $I$ transmitted through the soliton changes in time with the same frequency as the frequency $f$ of the AC field, Figs. 1b-d, 2b-d. The intensity $I$ depends on $\varphi_m$. In Fig. 1b, 2b, counterclockwise rotation $\varphi_m > 0$ results in a higher $I$, while $\varphi_m < 0$ reduces $I$. By using Jones matrices [10], we calculate $I(x,y,t)$ as a function of $\varphi_m(x,y,t)$ and plot the director for a single period of the AC field in Figs. 1b, 2b. The azimuthal reorientations are weak, reaching a maximum $\varphi_{max} \approx 5°$, for both $B_0^1$ and $B_{90}^1$. For clarity, the tilts of the director in Figs. 1b, 2b are enlarged by a factor of 6.

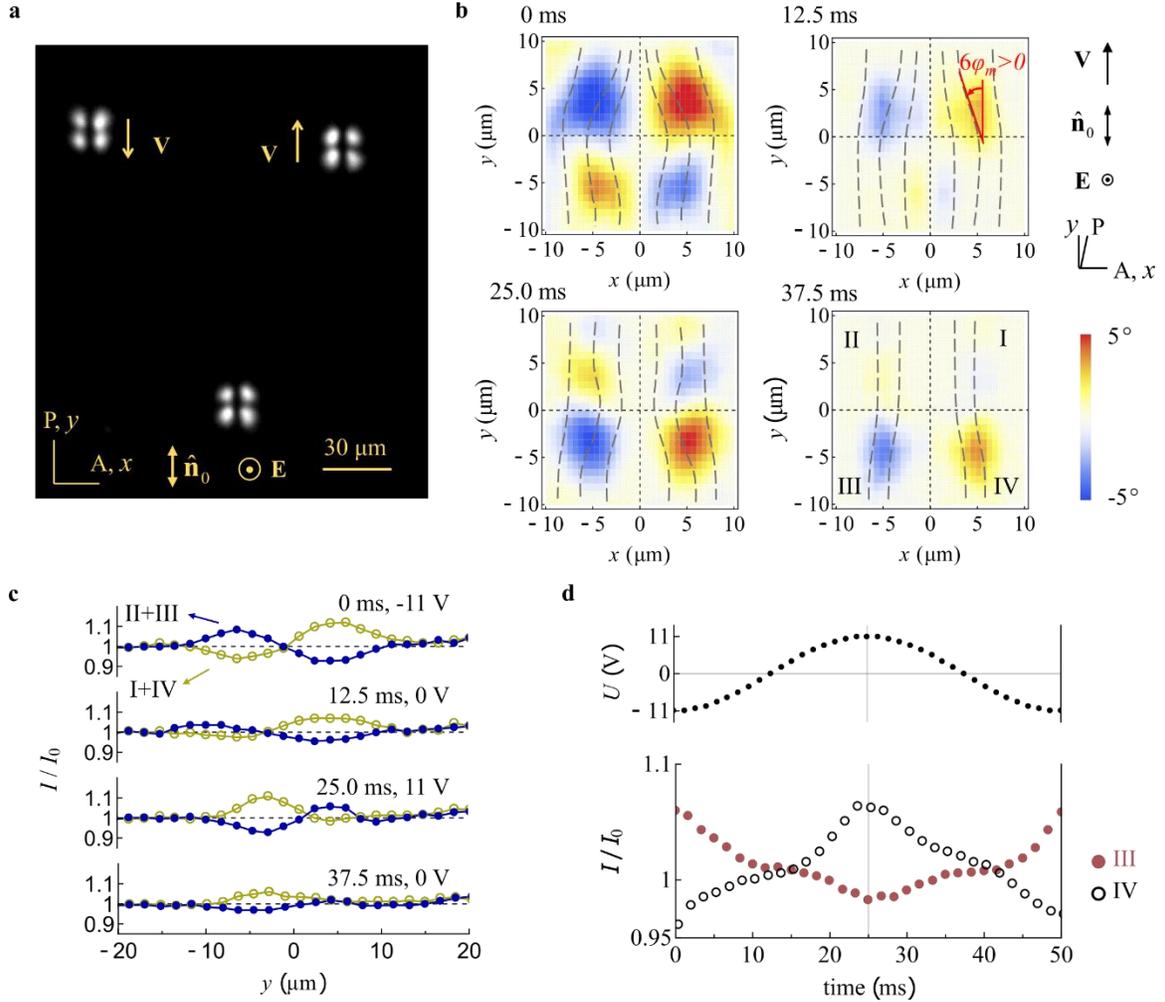

**Fig. 1. $B_0^1$ solitons (20 Hz, $U$=8 V). a,** Polarizing microscope texture. **b,** Azimuthal director distortions $\varphi_m(x,y,t)$ reproduced from polarizing microscopy with two polarizers crossed at $78°$. The time step between images is 1/4 of the voltage period, with "0 ms" corresponding to the negative extremum of the voltage; $\sqrt{2}\,U \approx 11.3$ V. **c,** $y$-profile of the transmitted light intensity integrated over the $x$-axis in right (I+IV) and left (II+III) parts of the bullet at different voltages/times. **d,** Time/voltage dependence of light intensity transmitted through quadrants III and IV. In **c**, **d**, $I$ is normalized by the transmitted intensity $I_0$ outside the bullet.

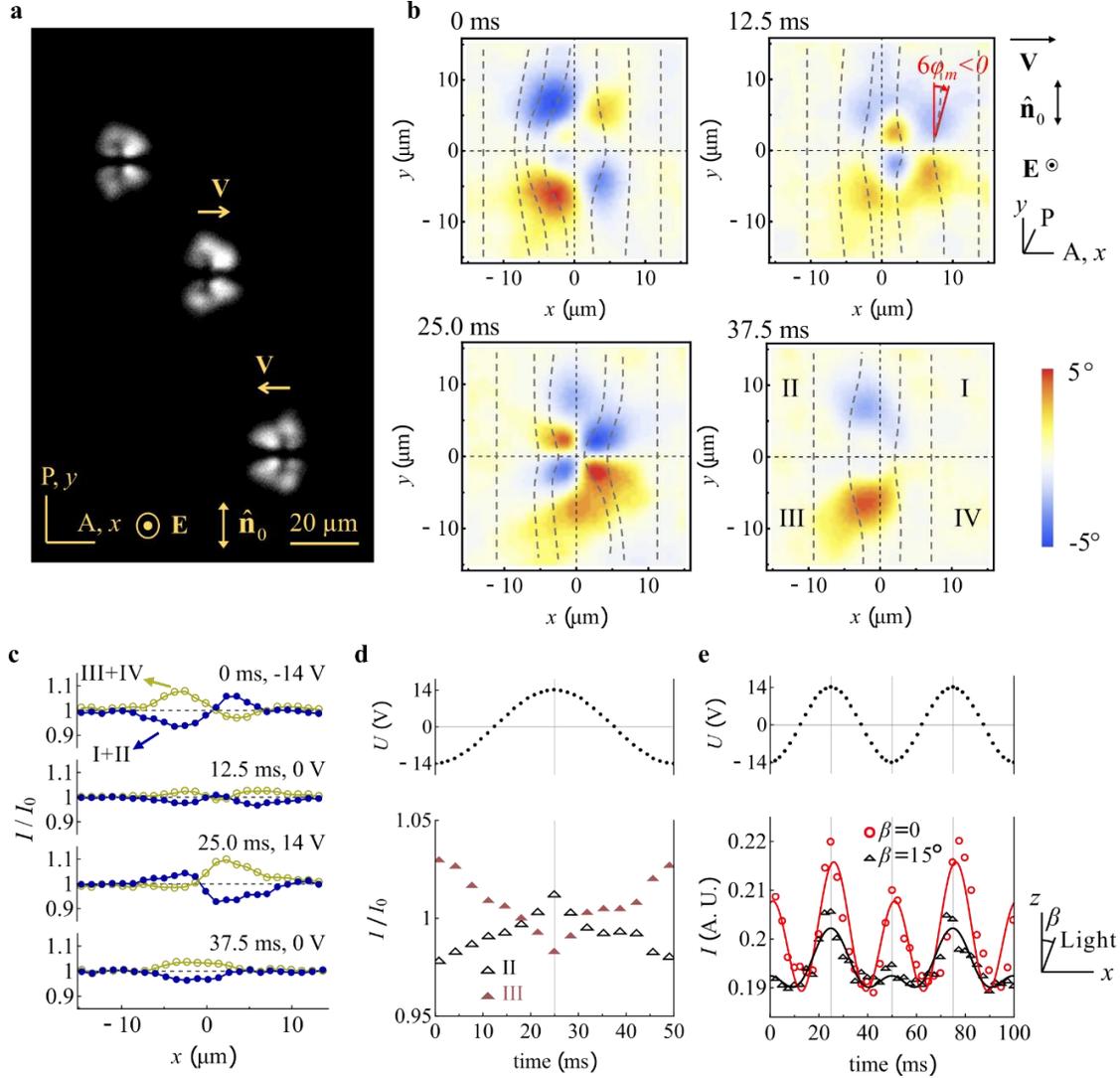

**Fig. 2.** $B_{90}^1$ **solitons (20 Hz, $U=10$ V). a,** Polarizing microscope texture. **b,** Azimuthal director distortions $\varphi_m(x,y,t)$ reproduced with two polarizers crossed at $65°$. The time step between images is 1/4 of the voltage period, time "0" corresponds to the negative extremum of the voltage; $\sqrt{2}\,U \approx 14.1$ V. **c,** $x$-profile of the light intensity integrated over the $y$-axis in top (I+II) and bottom (III+IV) parts of the bullet at different voltages/times. **d,** Time/voltage dependence of light intensity transmitted through quadrants II and III. **e,** Dynamics of light intensity at the quadrant I for the crossed polarizers at normal, $\beta = 0$, and oblique incidence, $\beta = 15°$. The solid lines show the fitting described in Supplementary text. In **c**, **d**, $I$ is normalized by the transmitted intensity $I_0$ outside the bullet.

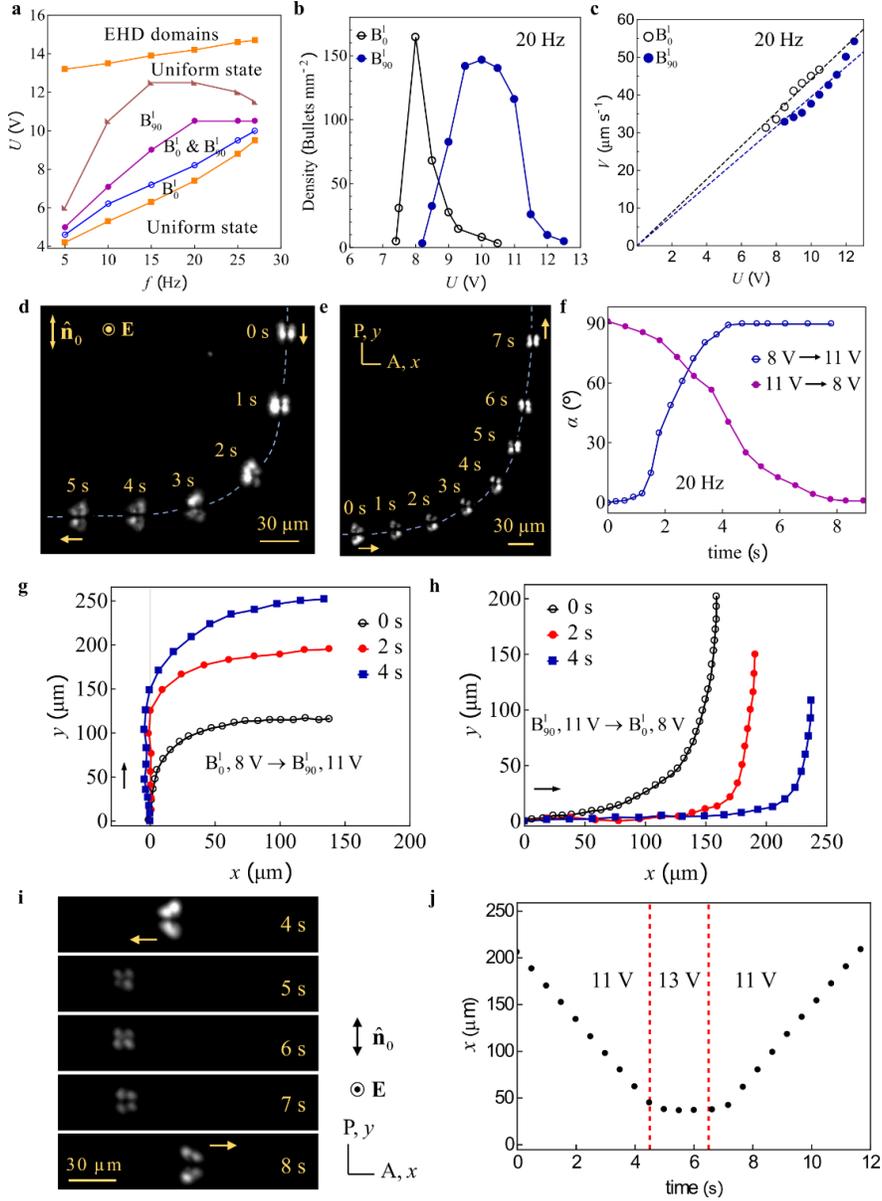

**Fig. 3. Soliton properties as a function of the applied voltage. a,** Phase diagram of solitons at $55°C$. Voltage dependence of **b** densities and **c** velocities of $B_0^l$ and $B_{90}^l$. **d,** $B_0^l$ transforms into $B_{90}^l$ when the voltage is raised at $t = 0$ s from 8 V to 11 V; **e,** The soliton $B_{90}^l$ transform into $B_0^l$ when the voltage at $t = 0$ s is decreased from 11 V to 8 V. **f,** Time dependence of the angle $\alpha$ between the propagation direction and $\hat{n}_0$ for transformations

shown in **d,e**. **g,** Changes of propagation directions during $B_0^1 \to B_{90}^1$ and **h,** $B_{90}^1 \to B_0^1$ transformation for different rates of the voltage change from 8 V to 11 V and back; the voltage is ramped linearly within 0 s, 2 s, and 4 s. **i,** Velocity reversal of $B_{90}^1$ soliton when the voltage is switched from 11 V to 13 V and then back to 11 V. **j,** Dynamics of the $y$-location of the soliton $B_{90}^1$ shown in **i**.

The director also experiences oscillations of the polar angle $\theta_m(x,y,t)$ measured with respect to the $xy$ plane, with the same frequency $f$, Fig. 2e, as explained below. The theoretical model in Supplementary text predicts that for $B_{90}^1$ solitons, when the incidence plane is perpendicular to $\hat{\mathbf{n}}_0$, the transmitted intensity depends on the angle of incidence $\beta$, measured between the probing beam and the normal to the cell, and the azimuthal $\varphi_m$ and polar $\theta_m$ director angles in the middle plane,

$$I \propto I_{\text{leak}} + \frac{4\Gamma^2}{\pi^2}\left(1-0.04\Gamma^2\right)\left(\varphi_m - \theta_m \tan\beta_{\text{LC}}\right)^2, \qquad (1)$$

where $I_{\text{leak}}$ is the light intensity leaked while the beam propagates between the crossed polarizers and the cell, normalized by the incident intensity, $\Gamma = 2\pi\Delta n\, d/\lambda \approx 1.7$ is the phase retardation of the undistorted nematic, determined by the wavelength $\lambda \approx 530$ nm and birefringence $\Delta n = n_e - n_o$, here $n_o$ and $n_e$ are the ordinary and extraordinary refractive indices, respectively; $\beta_{\text{LC}} = \beta/n_o$ is the angle between the probing beam and the normal to the cell within the nematic. Eq. (1) simplifies for normal incidence, $\beta_{\text{LC}} = \beta = 0$, as $I$ depends only on $\varphi_m$. The experiment, Fig. 2e, demonstrates that the transmission peaks for the two half-periods of the applied field are different. Since $I \propto \varphi_m^2$, the angle $\varphi_m$ changes with the frequency $f$ and has a non-zero stationary value. For oblique incidence, $\beta_{LC} = 10.6°$, the experiment and modeling, Fig.2e, show that $\theta_m$ also oscillates with the same frequency $f$. Figure 2e and Eq. (1) demonstrate that the variations of $\theta_m$ in the I quadrant are in phase with the variations of $\varphi_m$, as the term $\theta_m \tan \beta_{LC}$ diminishes the overall intensity of the transmitted light. Finally, numerical fitting using Eq. (1) allows us to estimate $\theta_m \approx 10° - 12°$.

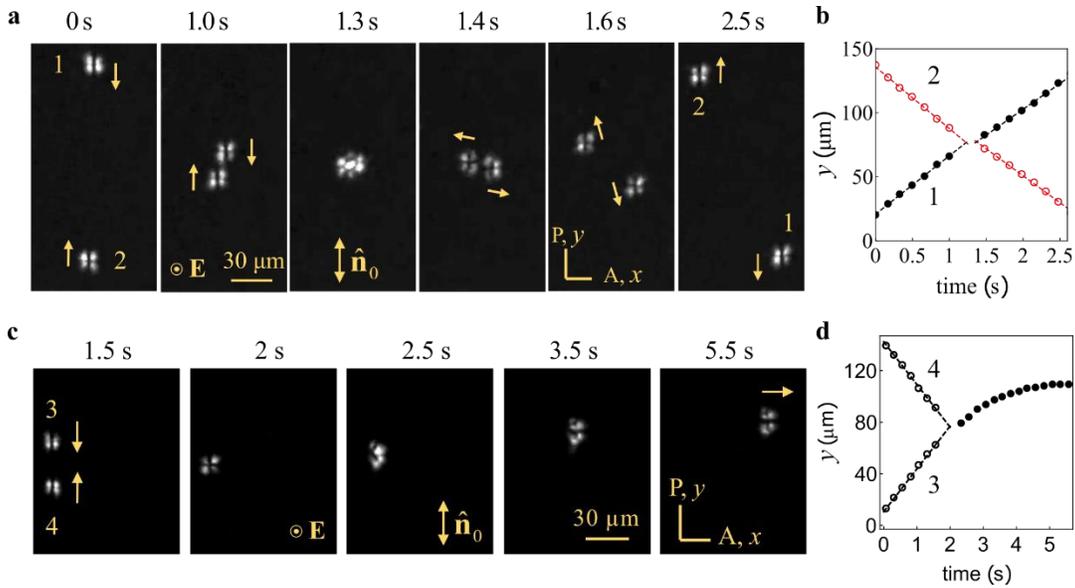

**Fig. 4. Collisions of $B_0^1$ solitons ; $f = 20$ Hz. a,** Collision of two $B_0^1$ solitons ("1" and "2") . **b,** Dynamics of the *y*-locations of solitons shown in **a**. **c,** Two $B_0^1$ solitons ("3" and "4") collide and transform into a single $B_{90}^1$ soliton (20 Hz, 8.5 V). **d,** Dynamics of the *y*-locations of the solitons shown in **c**.

The $B_\alpha^1$ solitons are sensitive to the anisotropy of dielectric permittivity, electric conductivity, concentration of ions, frequency and amplitude of the field. For example, in CCN-47 kept at 55°C , they exist in a narrow range of conductivities, $\sigma_\perp \approx (1.2 \div 2.5) \times 10^{-8} \, \Omega^{-1} \text{m}^{-1}$. In this range, as $\sigma_\perp$ increases, the range of bullets stability shifts towards higher frequencies. The stability region of $B_\alpha^1$ solitons corresponds to the so-called "conductivity" regime limited from above by a critical frequency $f_c$ introduced by Dubois-Violette et al [11,12] to describe electrohydrodynamics of LCs, see the Supplementary Text. Since the director oscillates with the same frequency as the field, the main reason of soliton existence is flexoelectric polarization and space charge developed at director deformations; both of these couple linearly to the field.

As illustrated by Fig.1b and 2b, the time-average director asymmetry develops parallel to $\hat{\mathbf{n}}_0$ in $B_0^1$ and perpendicular to $\hat{\mathbf{n}}_0$ in $B_{90}^1$ case; the change in the field amplitude controls the type of asymmetry and thus the prevailing direction of propagation. Solitons that propagate under the angle $0 < \alpha < 90°$ show no mirror plane symmetry, Fig. 3d,e.

In summary, we demonstrate electrically driven multidimensional (3+2)D solitons in a uniform nematic that can be controllably steered along different directions perpendicular to the applied field. The solitons represent perturbation of the director that are topologically trivial but self-confined along the longitudinal and transverse directions. The azimuthal and polar tilts of the director oscillate with the same frequency as the driving AC electric field. The speed and propagation direction of the solitons is controlled by the field. When a soliton propagates along or perpendicularly to $\hat{\mathbf{n}}_0$, its dynamic director structure shows the mirror symmetry plane that contains the propagation direction. For intermediate propagation directions, $0 < \alpha < 90°$, the solitons show no mirror symmetry. The ability of the solitons to move along different directions opens broad field of studies of their nontrivial interactions and collisions. It also promises practical applications.

Controlling the in-plane dynamics of the solitons, one can develop devices for targeted 2D delivery of optical information. Furthermore, since the solitons represent director deformations and since director deformations attract colloidal particles [13], the (3+2)D solitons can be used as micro-cargo vehicles.

**Materials and Methods**

**Materials.** We use a single-component nematic LC, 4'-butyl-4-heptyl-bicyclohexyl-4-carbonitrile, abbreviated as CCN-47 (Nematel GmbH). The conductivity was adjusted by adding 0.005wt% tetrabutylammonium bromide (TBABr, produced by Sigma-Aldrich) using chloroform as a solvent. Chloroform was evaporated in the vacuum oven for 24 hours at the room temperature and then for another 24 hours at $60\,°C$. We measured the anisotropy of both permittivity and conductivity by an LCR meter 4284A (Hewlett-Packard) using cells with planar and homeotropic alignment using polyimides AL1254 and JALS-204, respectively (both are purchased from Japan Synthetic Rubber Co.). The principal components of conductivity and dielectric permittivity tensors of doped CCN-47 are $\sigma_{\parallel} \approx 1.6\times10^{-8}\,\Omega^{-1}\mathrm{m}^{-1}$, $\sigma_{\perp} \approx 1.8\times10^{-8}\,\Omega^{-1}\mathrm{m}^{-1}$, $\varepsilon_{\parallel} \approx 4.9$, and $\varepsilon_{\perp} \approx 8.2$ at 5 kHz and $55°C$; the subscripts indicate whether the property is measured along the background director $\hat{\mathbf{n}}_0$ or perpendicularly to it.

**Generation of solitons.** The cell is composed of two glass substrates coated with indium tin oxide (ITO), which serve as the transparent electrode of active area $5\times5\,\mathrm{mm}^2$. The alignment layers PI-2555 coated on the surface of ITO were rubbed to provide a planar alignment. The temperature of the cell is controlled with a Linkam LTS350 hot stage and a Linkam TMS94 controller. The AC voltage is applied using a waveform generator (Stanford Research Systems, Model DS345) and an amplifier (Krohn-hite Corporation, Model 7602).

**Optical characterization of solitons.** We use a polarizing Nikon TE2000 microscope equipped with two cameras: Emergent HR20000 with the maximum frame rate 1000 fps and MotionBLITZ EOSens mini1 (Mikrotron GmbH) with the maximum frame rate 6000 fps. The diagram in Fig. 3A was established by performing voltage scan with 0.1 V increments at $f = const$; at each voltage level, the system was stabilized for 5 min before measurements. The location of soliton was tracked by an open-source software ImageJ and its plugin TrackMate. Measuring the $x$, $y$ coordinates of the solitons as a function of time yields the velocity. The azimuthal distortion of the director deviating from the background $\hat{\mathbf{n}}_0$ was determined by numerical simulations of the transmitted light intensity based on Jones matrix, in the geometry with two linear polarizers crossed at $78°$ and $65°$. The nematic slab was split into thin layers, with the director twist $\varphi(i) = \varphi_m \sin\left(\frac{2i-1}{400}\pi\right)$, where $i$ is an integer in the range from 1 to 200 and $\varphi_m = \varphi_m(x, y, t)$ is the azimuthal distortion of the director in the middle of cell, $i = 100$. Using the measured birefringence of CCN-47 $\Delta n = 0.018$ at $55°C$, cell thickness $d = 8\,\mu\text{m}$, wavelength of light $\lambda \approx 530\,\text{nm}$, we calculate the ratio $T_{\text{bullet}}/T_0$ of the light intensity transmitted through the soliton to the light intensity transmitted through the background region, as a function of $\varphi_m$. Since $T_{\text{bullet}}/T_0$ equals the experimentally determined ratio $I_{\text{bullet}}/I_0$, where $I$ is the transmitted light intensity through the bullet and $I_0$ is the intensity of light transmitted through the background uniform region, the dependency $I_{\text{bullet}}/I_0$ on $\varphi_m$ allows us to map $\varphi_m(x, y, t)$ in Fig. 1B and 2B. The symmetry of director distortion of solitons is additionally verified by using complementary angles of polarizers' decrossing, such as 102° and 115°.

**References**

1   Kartashov, Y. V., Astrakharchik, G. E., Malomed, B. A. & Torner, L. Frontiers in multidimensional self-trapping of nonlinear fields and matter. *Nat. Rev. Phys.* **1**, 185-197 (2019).
2   Dauxois, T. & Peyrard, M. *Physics of solitons*. (Cambridge University Press, 2006).
3   Chen, Z., Segev, M. & Christodoulides, D. N. Optical spatial solitons: historical overview and recent advances. *Rep. Prog. Phys.* **75**, 086401 (2012).
4   Peccianti, M. & Assanto, G. Nematicons. *Phys. Rep.* **516**, 147-208 (2012).
5   Hang, C. & Huang, G. Guiding ultraslow weak-light bullets with Airy beams in a coherent atomic system. *Phys. Rev. A* **89**, 013821 (2014).
6   Malomed, B. A. Multidimensional solitons: Well-established results and novel findings. *Eur. Phys. J. Spec. Top.* **225**, 2507-2532 (2016).
7   Gan, X., Zhang, P., Liu, S., Xiao, F. & Zhao, J. Beam steering and topological transformations driven by interactions between a discrete vortex soliton and a discrete fundamental soliton. *Phys. Rev. A* **89**, 013844 (2014).
8   Sala, F. A. *et al.* Bending reorientational solitons with modulated alignment. *J. Opt. Soc. Am. B* **34**, 2459-2466 (2017).
9   Li, B.-X. *et al.* Electrically driven three-dimensional solitary waves as director bullets in nematic liquid crystals. *Nat. Commun.* **9**, 2912 (2018).
10  Jones, R. C. A new calculus for the treatment of optical systemsI. description and discussion of the calculus. *J. Opt. Soc. Am.* **31**, 488-493 (1941).
11  de Gennes, P. G. & Prost, J. *The Physics of Liquid Crystals*. (Clarendon Press, 1995).
12  Dubois-Violette, E., de Gennes, P. G. & Parodi, O. Hydrodynamic instabilities of nematic liquid crystals under A. C. electric fields. *J. Phys. France* **32**, 305-317 (1971).

13      Voloschenko, D., Pishnyak, O. P., Shiyanovskii, S. V. & Lavrentovich, O. D. Effect of director distortions on morphologies of phase separation in liquid crystals. *Phys. Rev. E* **65**, 060701 (2002).

14      Berreman, D. W. Optics in Stratified and Anisotropic Media - 4x4-Matrix Formulation. *J. Opt. Soc. Am.* **62**, 502-510 (1972).

15      Allia, P., Oldano, C. & Trossi, L. Polarization Transfer-Matrix for the Transmission of Light through Liquid-Crystal Slabs. *Journal of the Optical Society of America B-Optical Physics* **5**, 2452-2461 (1988).

16      Kiselev, A. D., Vovk, R. G., Egorov, R. I. & Chigrinov, V. G. Polarization-resolved angular patterns of nematic liquid crystal cells: Topological events driven by incident light polarization. *Phys. Rev. A* **78**, 033815 (2008).

Reprints and permissions information is available at www.nature.com/reprints

**Acknowledgements**

The work was supported by NSF grants DMS-1729509.

**Author Contributions:**

BL discovered the studied solitons. SVS developed the model of light transmittance through the soliton. BL and RX performed the experimental studies. SP measured the conductivity of the studied materials. BL, RX, SVS, and ODL analyzed the data and discussed the results. ODL directed the research and wrote the manuscript with an input from all coauthors.

**Author Information:**

Correspondence and requests for materials should be addressed to olavrent@kent.edu.

**Competing interests:**

The authors declare no competing interests.

**Supplemental Information**

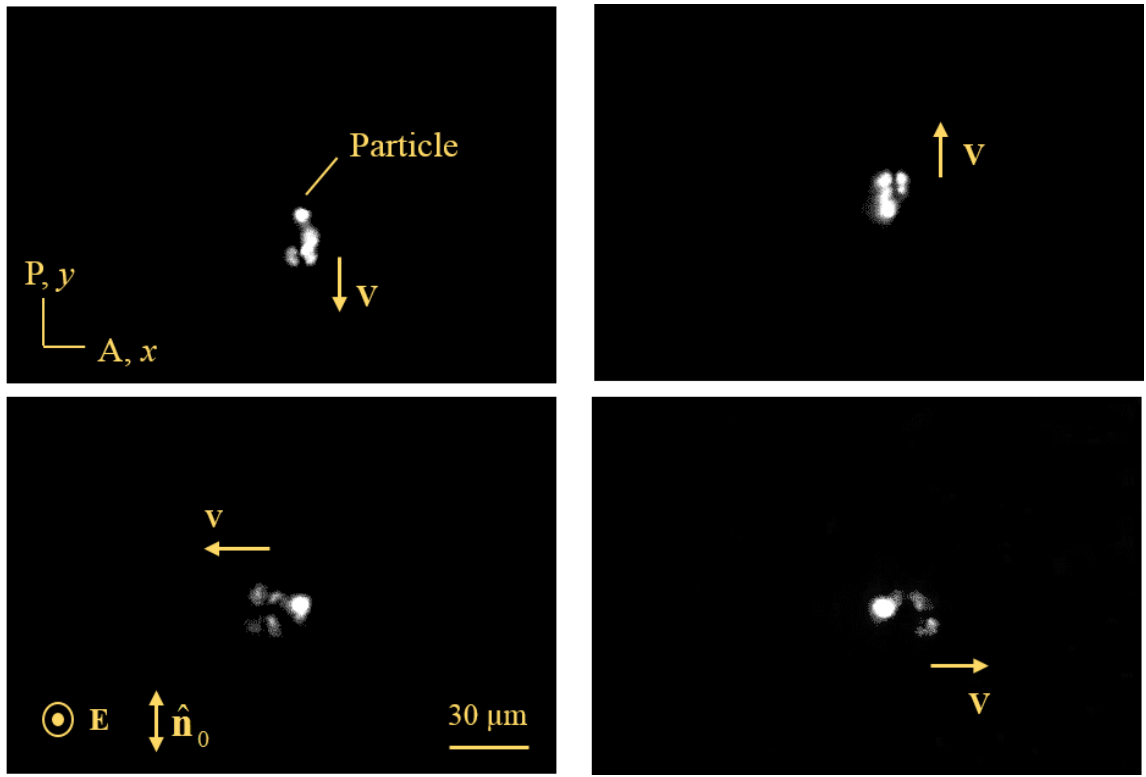

**Supplementary Fig. 1| Creation of solitons $B_0^1$ and $B_{90}^1$ from the same dust particle at 55 °C (20 Hz, 9 V).**

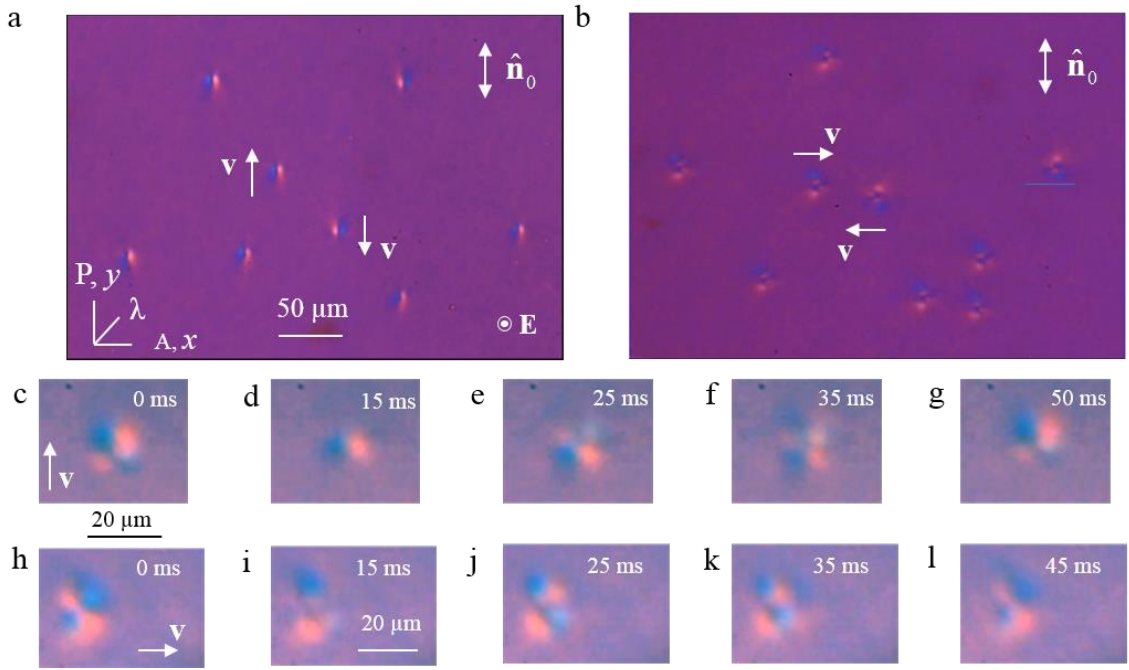

**Supplementary Fig. 2|  Solitons are observed between crossed polarizers and a compensator with the optic axis λ. a,** Solitons $B_0^1$ propagate along the director $\hat{\mathbf{n}}_0$ (20 Hz, 8 V); **b,** Solitons $B_{90}^1$ propagate perpendicular to $\hat{\mathbf{n}}_0$ (20 Hz, 11.0 V). The electric field is normal to the *xy* plane. **c-g,** time sequence of polarizing microscope images of the $B_0^1$ soliton (20 Hz, 8 V). **h-l,** the same for the $B_{90}^1$ soliton (20 Hz, 11.0 V).

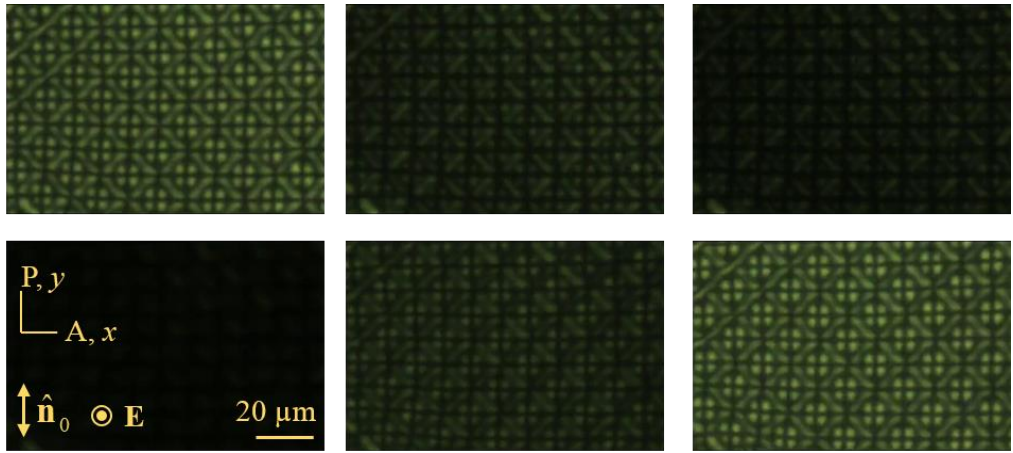

**Supplementary Fig. 3| Sequence of cross-type electro-convection patterns in half period of applied voltage.** The images are taken with crossed polarizers (20 Hz, 14.7V).

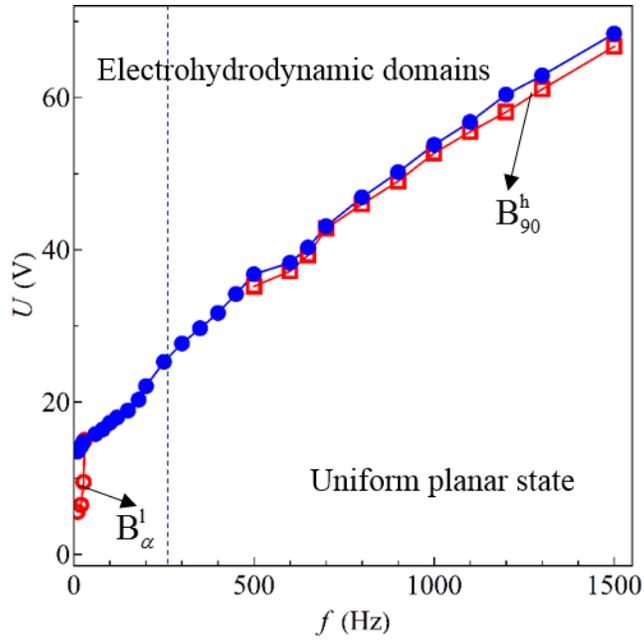

**Supplementary Fig. 4| Phase diagram of director bullets in the studied material at 55 ºC.** Symbol $B_{90}^h$ denotes (3+1)D director bullets that form in the applied electric field of a high frequency; these bullets can propagate only perpendicularly to the director, as described in Ref. (9).

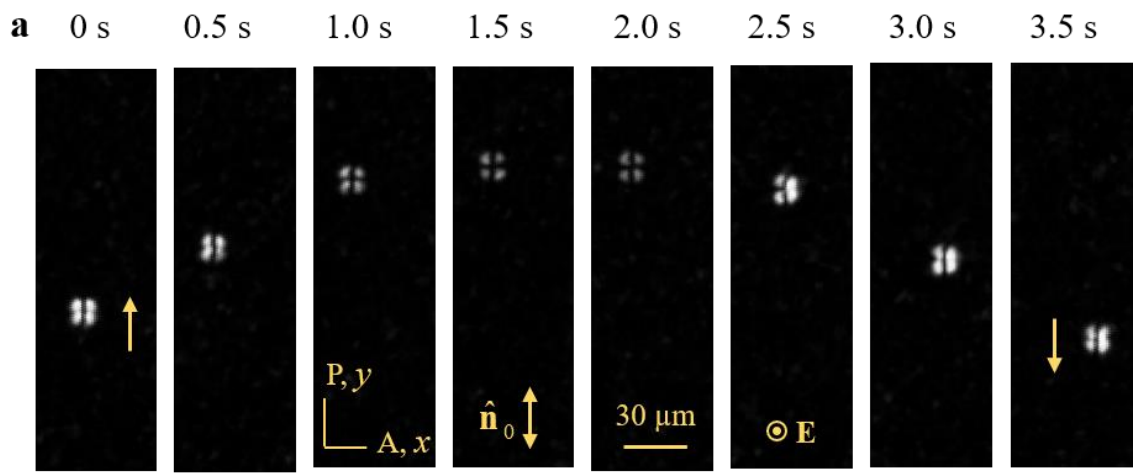

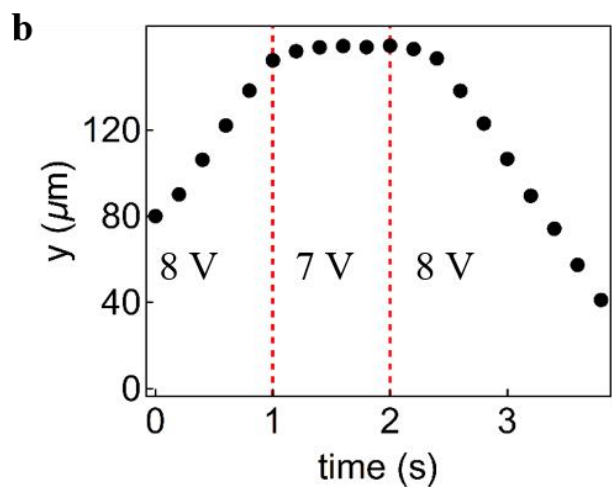

**Supplementary Fig. 5| Velocity reversal of a soliton $B_0^1$ by switching the voltage from 8 V to 7 V and then back to 8 V. a,** Time sequence of the microscopy textures under crossed polarizers. **b**, Time dependence of the *y*-coordinates of the soliton.

1. **Light propagation through nematic bullets: Theory**

We consider electromagnetic wave propagation in one-dimensionally distorted uniaxial nematic within Berreman's 4x4 matrix formalism [14] in the representation similar to Refs.[15,16]. For a monochromatic wave with the electric $\mathbf{E}$ and magnetic $\mathbf{H}$ fields changing as $\propto e^{-i\omega t}$, we use the dimensionless coordinates $\tilde{\mathbf{r}} = \kappa\mathbf{r}$, where $\kappa = \omega/c = 2\pi/\lambda$ is the free space wavenumber. We assume that the dielectric tensor at optical frequencies (optic tensor) $\boldsymbol{\varepsilon}$ with the elements $\varepsilon_{ij}$ depends only on the coordinate $z$ perpendicular to the nematic slab, and does not depend on the in-plane coordinates $\boldsymbol{\rho} = \{\tilde{x}, \tilde{y}\}$; here and in what follows, tilde implies dimensionless coordinates, as specified above. The assumption preserves the in-plane 2D wave vector $\mathbf{q} = (q_1, q_2)$ and allows us to obtain the solution of the Maxwell equations in the form of 4-vector $\tilde{z}$-dependent amplitudes $(E_x, E_y, H_y, -H_x)\exp(i\mathbf{q}\boldsymbol{\rho})$, where $\mathbf{F} = (E_x, E_y, H_y, -H_x)$ obeys the equation:

$$\mathbf{F}' = i\mathbf{M}\cdot\mathbf{F}, \qquad (1)$$

where $' = \partial/\partial\tilde{z}$ and the 4x4 matrix $\mathbf{M}$ is:

$$\mathbf{M} = \begin{pmatrix} -q_1\varepsilon_{31}/\varepsilon_{33} & -q_1\varepsilon_{32}/\varepsilon_{33} & 1-(q_1^2/\varepsilon_{33}) & -q_1q_2/\varepsilon_{33} \\ -q_2\varepsilon_{31}/\varepsilon_{33} & -q_2\varepsilon_{32}/\varepsilon_{33} & -q_1q_2/\varepsilon_{33} & 1-(q_2^2/\varepsilon_{33}) \\ \varepsilon_{11}-q_1^2-(\varepsilon_{13}\varepsilon_{31}/\varepsilon_{33}) & \varepsilon_{12}+q_1q_2-(\varepsilon_{13}\varepsilon_{32}/\varepsilon_{33}) & -q_1\varepsilon_{13}/\varepsilon_{33} & -q_2\varepsilon_{13}/\varepsilon_{33} \\ \varepsilon_{21}+q_1q_2-(\varepsilon_{23}\varepsilon_{31}/\varepsilon_{33}) & \varepsilon_{22}-q_2^2-(\varepsilon_{23}\varepsilon_{32}/\varepsilon_{33}) & -q_1\varepsilon_{23}/\varepsilon_{33} & -q_2\varepsilon_{23}/\varepsilon_{33} \end{pmatrix}.$$

(2)

In a homogeneous medium, where $\boldsymbol{\varepsilon} = const$ and therefore $\mathbf{M} = \bar{\mathbf{M}} = const$ are $\tilde{z}$-independent, the general solution of (1) is

$$\bar{\mathbf{F}}(\tilde{z}) = \bar{\mathbf{V}} \cdot \bar{\mathbf{L}}(\tilde{z}) \cdot \bar{\mathbf{A}}, \tag{3}$$

where the matrices $\bar{\mathbf{V}} = \{\mathbf{V}_1, \mathbf{V}_2, \mathbf{V}_3, \mathbf{V}_4\}$ and $\bar{L}_{ij}(\tilde{z}) = \delta_{ij} \operatorname{Exp}(i\bar{k}_i \tilde{z})$ with $\alpha, \beta = 1, 2, 3, 4$ are built by the eigenvectors $\bar{\mathbf{V}}_i$ and the eigenvalues $\bar{k}_i$ of matrix $\bar{\mathbf{M}}$, that obey the equation

$$\bar{\mathbf{M}} \cdot \bar{\mathbf{V}}_i = \bar{k}_i \bar{\mathbf{V}}_i, \tag{4}$$

and the 4-vector $\mathbf{A}$ contains the constant amplitudes of the eigenwaves that are determined by the boundary conditions.

When the uniaxial LC is distorted in the body of the soliton, $\hat{\mathbf{n}} = \hat{\mathbf{n}}(\tilde{z})$, the associated inhomogeneity of the optic tensor $\boldsymbol{\varepsilon}(\tilde{z}) = n_o^2 \mathbf{I} + (n_e^2 - n_o^2)(\hat{\mathbf{n}}(\tilde{z}) \otimes \hat{\mathbf{n}}(\tilde{z}))$ is controlled by the birefringence $\Delta n = n_e - n_o$, where $n_o$ and $n_e$ are the ordinary and extraordinary refractive indices, respectively. Because in our experiments the azimuthal distortions are small and obey the condition $|\varphi| < \Gamma^{-1}$, where $\Gamma = 2\pi \Delta n \, d / \lambda \approx 1.7$ is the phase retardation of the undistorted LC area, we use the perturbation method to determine the transformation between ordinary and extraordinary waves.

We split $\mathbf{M}(\tilde{z}) = \bar{\mathbf{M}} + \tilde{\mathbf{M}}(\tilde{z})$ into a homogeneous part $\bar{\mathbf{M}}$ and an inhomogeneous $\tilde{\mathbf{M}}(\tilde{z})$ part. The solution of (1) for this case can be presented as:

$$\mathbf{F}(\tilde{z}) = \bar{\mathbf{V}} \cdot \bar{\mathbf{L}}(\tilde{z}) \cdot \mathbf{A}(\tilde{z}), \tag{5}$$

where the matrices $\bar{\mathbf{V}}$ and $\bar{L}_{ij}(\tilde{z}) = \delta_{ij} \operatorname{Exp}(i\bar{k}_i \tilde{z})$ correspond to the homogeneous part and obey Eq.(4). Substituting this solution into Eq.(1), we obtain the equation that defines evolution of the eigenwaves' amplitudes 4-vector $\mathbf{A}(\tilde{z})$ along the $z$-axis

$$\mathbf{A}'(\tilde{z}) = \bar{\mathbf{L}}^{-1}(\tilde{z}) \cdot \tilde{\mathbf{S}}(\tilde{z}) \cdot \bar{\mathbf{L}}(\tilde{z}) \cdot \mathbf{A}(\tilde{z}), \tag{6}$$

where $\tilde{\mathbf{S}}(\tilde{z}) = i\,\bar{\mathbf{V}}^{-1} \cdot \tilde{\mathbf{M}}(\tilde{z}) \cdot \bar{\mathbf{V}}$ is the matrix, which changes the amplitudes of eigenwaves of $\bar{\mathbf{M}}$. The linearity of Eq.(6) allows us to separate boundary conditions from evolution of the amplitudes of eigenmodes in the bulk of the inhomogeneous medium, by introducing the propagation matrix $\mathbf{U}(\tilde{z})$,

$$\mathbf{A}(\tilde{z}) = \mathbf{U}(\tilde{z}) \cdot \mathbf{A}(0). \tag{7}$$

The propagation matrix $\mathbf{U}(\tilde{z})$ obeys the boundary condition $\mathbf{U}(0) = \mathbf{I}_4$, where $\mathbf{I}_4$ is the 4×4 unit matrix) and the equation similar to Eq. (6),

$$\mathbf{U}'(\tilde{z}) = \bar{\mathbf{L}}^{-1}(\tilde{z}) \cdot \tilde{\mathbf{S}}(\tilde{z}) \cdot \bar{\mathbf{L}}(\tilde{z}) \cdot \mathbf{U}(\tilde{z}). \tag{8}$$

The solution of the last equation in the approximation linear on $\tilde{\mathbf{S}}(\tilde{z})$,

$$U_{ij}(\tilde{z}) = \delta_{ij} + \int_0^{\tilde{z}} \tilde{S}_{ij}(\bar{z}) \operatorname{Exp}\left[i\left(\bar{k}_j - \bar{k}_i\right)\bar{z}\right] d\bar{z}, \tag{9}$$

is valid if the integral contributions are smaller than 1.

Because of the strong planar anchoring at the bounding plates, $\hat{\mathbf{n}}(0) = \hat{\mathbf{n}}(\tilde{d}) = \hat{\mathbf{n}}_0 = (0,1,0)$, we select $\bar{\mathbf{M}}$ that corresponds to $\hat{\mathbf{n}}_0$ and assume that the polar angle $\theta(\tilde{z}) = \theta_m \sin(\pi \tilde{z}/\tilde{d})$ and the azimuthal angle $\varphi(\tilde{z}) = \varphi_m \sin(\pi \tilde{z}/\tilde{d})$, which define the director field $\hat{\mathbf{n}} = (\cos\theta\sin\varphi, \cos\theta\cos\varphi, \sin\theta)$ have simple one harmonic z-dependence with small amplitudes $\theta_m$ and $\varphi_m$. The transmitted light intensity in the optical scheme with two crossed linear polarizers is determined by $|U_{12}(\tilde{d})|^2$. The $xz$ is the incidence plane, $\left|U_{12}(\tilde{d})\right|^2 = \frac{4\Gamma^2}{\pi^2}[\varphi_m - \theta_m \tan\beta_{LC}]^2 G(\Gamma)$, where $G(\Gamma) \approx 1 - 0.04\Gamma^2$ is a slowly decaying function and $\beta_{LC}$ is the angle of propagation in the LC medium, measured with respect to the $z$-axis. When $yz$ is the incidence plane, then $\left|U_{12}(\Gamma)\right|^2 = \frac{2\Gamma^2}{\pi}\varphi_m^2\left[1 + (\frac{\pi}{4}\theta_m \tan\beta_{LC})^2\right]G(\Gamma)$. Taking into account the intensity of light leaked while the beam propagates between the crossed polarizers $I_{leak}$ we result in the measured intensity as

$$I \approx I_{leak} + \left|U_{12}(\tilde{d})\right|^2. \qquad (10)$$

2. **Conductive and dielectric regimes of the studied nematics**

Unlike (3+1)D bullets describe previously for relatively high frequencies $f > 100$ Hz, $B_\alpha^1$ of the (3+2)D type exist at frequencies 30 Hz and lower. This region can be associated with the so-called "conductivity" regime limited from above by the critical frequency $f_c = \frac{1}{\tau_M}\sqrt{\xi^2 - 1}$ which is used to describe electrohydrodynamics of LCs. Here $\tau_M = \frac{\sigma_\perp}{\varepsilon_0 \varepsilon_\perp}$ is the Maxwell relaxation time for a planar cell, $\varepsilon_0 = 8.85 \times 10^{-12}$ F/m, and

$\xi^2 = \left(1 - \frac{\sigma_\perp}{\sigma_\parallel}\frac{\varepsilon_\parallel}{\varepsilon_\perp}\right)\left(1 + \frac{\alpha_2}{\eta_c}\frac{\varepsilon_\parallel}{\Delta\varepsilon}\right)$ is the material parameter that depends on conductivities $\sigma_\parallel$ and $\sigma_\perp$, permittivities. $\varepsilon_\parallel$ and $\varepsilon_\perp$, viscous coefficients $\alpha_2$ and $\eta_c$. Using the experimental data measured at 5 kHz and 55°C for the TBABr-doped CCN-47, we find $1/\tau_M \sim 248$ Hz. The factor $\sqrt{\xi^2 - 1}$ in the expression for $f_c$ is hard to determine exactly, since $\alpha_2$ and $\eta_c$ are not known. Using the independently measured parameters of TBABr-doped CCN-47, we find $\left(1-\frac{\sigma_\perp}{\sigma_\parallel}\frac{\varepsilon_\parallel}{\varepsilon_\perp}\right) \approx 0.3$ and $\frac{\varepsilon_\parallel}{\Delta\varepsilon} = -1.5$. With the ratio $\frac{-\alpha_2}{\eta_c}$ being on the order of 1, one finds that $\sqrt{\xi^2 - 1}$ is on the order of 1 or even smaller. This result implies that the dynamic solitons that appear in the frequency range below 30 Hz correspond to the conductive regime of LC electrohydrodynamics.